\documentclass[conference]{IEEEtran}
\IEEEoverridecommandlockouts
\usepackage{cite}
\usepackage{amsmath,amssymb,amsfonts}
\usepackage{algorithmic}
\usepackage{graphicx}
\usepackage{textcomp}
\usepackage{xcolor}
\usepackage{subfigure}
\usepackage{diagbox}

\def\BibTeX{{\rm B\kern-.05em{\sc i\kern-.025em b}\kern-.08em
    T\kern-.1667em\lower.7ex\hbox{E}\kern-.125emX}}
\begin{document}

\title{PCB Defect Detection Using Denoising Convolutional Autoencoders\\
}

\author{\IEEEauthorblockN{Saeed Khalilian}
\IEEEauthorblockA{\textit{Faculty of Computer Engineering} \\
\textit{University of Isfahan}\\
Isfahan, Iran \\
saeidkhalilian@eng.ui.ac.ir}
\and
\IEEEauthorblockN{Yeganeh Hallaj}
\IEEEauthorblockA{\textit{Faculty of Computer Engineering} \\
\textit{University of Isfahan}\\
Isfahan, Iran \\
y.hallaj@eng.ui.ac.ir}
\and
\IEEEauthorblockN{Arian Balouchestani}
\IEEEauthorblockA{\textit{Faculty of Computer Engineering} \\
\textit{University of Isfahan}\\
Isfahan, Iran \\
arian\_balouchestani@eng.ui.ac.ir}
\and
\IEEEauthorblockN{Hossein Karshenas}
\IEEEauthorblockA{\textit{Faculty of Computer Engineering} \\
\textit{University of Isfahan}\\
Isfahan, Iran \\
h.karshenas@eng.ui.ac.ir}
\and
\IEEEauthorblockN{Amir Mohammadi}
\IEEEauthorblockA{\textit{Industrial Automation Unit} \\
\textit{Pardisan Rayaneh System Company}\\
Isfahan, Iran \\
mohammadi@pardisan-co.com}
}

\maketitle

\begin{abstract}
Printed Circuit boards (PCBs) are one of the most important stages in making electronic products. A small defect in PCBs can cause significant flaws in the final product. Hence, detecting all defects in PCBs and locating them is essential. In this paper, we propose an approach based on denoising convolutional autoencoders for detecting defective PCBs and to locate the defects. Denoising autoencoders take a corrupted image and try to recover the intact image. We trained our model with defective PCBs and forced it to repair the defective parts. Our model not only detects all kinds of defects and locates them, but it can also repair them as well. By subtracting the repaired output from the input, the defective parts are located. The experimental results indicate that our model detects the defective PCBs with high accuracy (97.5\%) compare to state of the art works. 
\end{abstract}


\section{Introduction}
Printed Circuit Board (PCB) is a collection of electronic boards that helps different electronic components connect to each other. It is used in every electronic product and with its help, the number of errors in assembling is greatly reduced \cite{b1}. PCB is made of multiple sheet layers made of copper in which electronic components such as electrical resistances or capacitors are placed. Since PCBs are the first step in manufacturing an electronic device, a simple error in them leads to huge flaws in the final product \cite{b2}. Therefore, it is vital to detect defective PCBs not only to prevent further obstacles but also to fix the errors with as low cost as possible \cite{b3}. 

Visual inspection is the most important step in PCB manufacturing. PCB inspection consists of two major processes: defect detection and defect classification. In the past years, image processing techniques were widely used to automate these processes in PCB industries \cite{b4}. These techniques can locate defective parts in PCBs and classify them. However, due to the limitations and drawbacks of these techniques, they have been replaced with other methods such as machine learning methods in recent years \cite{b3}. 

Machine learning methods have attracted the attention of a variety of people from academia to industry. One of the most popular machine learning methods is neural networks, which consist of several nodes connected to each other to form a structure like the human brain. Convolutional neural network (CNN) is a special kind of neural networks that consists of several hidden layers. Autoencoder is an unsupervised neural network that tries to code inputs into a set of features and then decode them again to achieve outputs \cite{b5}. Autoencoder is useful for extracting different features from a data set. There are different types of Autoencoder such as denoising autoencoders or sparse autoencoders. Denoising autoencoders take corrupted images as input and try to extract features from the parts that are not noisy to generate flawless outputs. 

In this paper, we propose a method to detect defective PCBs, locate their flaws and repair them with the help of denoising convolutional autoencoders. The rest of the paper is organized as follows: Section 2 investigates the previous similar methods in PCB defect detection, the proposed method is described in section 3 and in section 4 the results of our work are explained.

\section{Literature Review}

PCB defect detection schemes are divided into 2 main categories: image processing techniques and machine learning methods.

Image processing techniques can be used to detect errors in PCBs and classify them. Dave et al \cite{b6} proposed a reasonable PCB inspection system that detects defects in bar PCBs using image processing. This method can recognize commons defects such as missing holes or open circuits.  Wu et al \cite{b2} developed an automated visual inception system for PCBs. This method subtracts a template PCB image from inspected images and uses an elimination process to locate defects in PCBs. To group all possible defects in PCBs, Kamalpreet et al \cite{b1} presented a method using MATLAB image processing operations. This method groups 14 possible defects into 5 groups. In order to classify the PCB defects, Putera et al \cite{b7} proposed a PCB defect detection and classification system using a morphological image segmentation algorithm and image processing theories. This system detects and classifies the defects on bar single layer PCBs. Ibrahim et al \cite{b8} presented a scheme to locate any defects on PCBs automatically using a wavelet-based image difference algorithm. This scheme is more efficient compared to previous traditional methods.

Some of the researches used machine learning methods to improve the accuracy and efficiency of previous image processing techniques. Srimani et al \cite{b9} proposed a hybrid approach to detect and classify defects in PCBs using soft computing techniques. This approach uses an adaptive genetic algorithm for feature selection and a neural network classifier. To tackle the problem of solder-balls occurrence in PCBs, Kusiak et al \cite{b10} developed a method that uses a data mining approach to identify the cause of these defects. Vafeiadis et al \cite{b11} proposed a framework to compare different machine learning classifiers such as support vector machine (SVM) for defect detection in PCBs. The SVM classifier had the best accuracy due to the results. 

Deep Learning is a machine learning approach for recognizing patterns and classifying them. It works best with unstructured data and unlabeled datasets compare to other machine learning methods. Therefore, it is an impeccable approach for PCB defect detection. Wei et al \cite{b3} proposed a defect classification algorithm based on a convolutional neural network (CNN). Adibhatla et al \cite{b12} presented a method to recognize defective PCBs. Their method is based on CNN and it classifies each PCB as either defective or intact PCBs. Zhang et al \cite{b13} developed an approach for PCB defect detection by learning deep discriminative features. They used a sliding window approach to locate the defects. To automatic defect verification, Deng et al \cite{b14} proposed Auto-VRS. Auto-VRS uses deep neural networks which verifies whether a defect is real or not. To tackle the problem of multi-label learning Zhang et al \cite{b15} proposed a model that uses a multi-task convolutional neural network. This model defines label learning as a binary classification task. Tang et al \cite{b16} proposed a model based on deep learning to detect PCB defects. They have provided a dataset called DeepPCB which contains 1500 image pairs of defective and intact PCBs. Mujeeb et al \cite{b17} proposed a deep learning-based approach to detect defects. This approach uses Autoencoders to extract discriminative features. 

\section{The Proposed Method}

In this section, a method for defect detection in PCBs is proposed. The proposed method not only detects defective boards and locates the possible defects, but also repairs the defective PCBs. The proposed method is based on denoising convolutional autoencoders, and it is a comprehensive method to detect all possible defects. Figure 1 shows an overview of the proposed method.

\begin{figure}[htbp]
\centerline{\includegraphics[width=\linewidth]{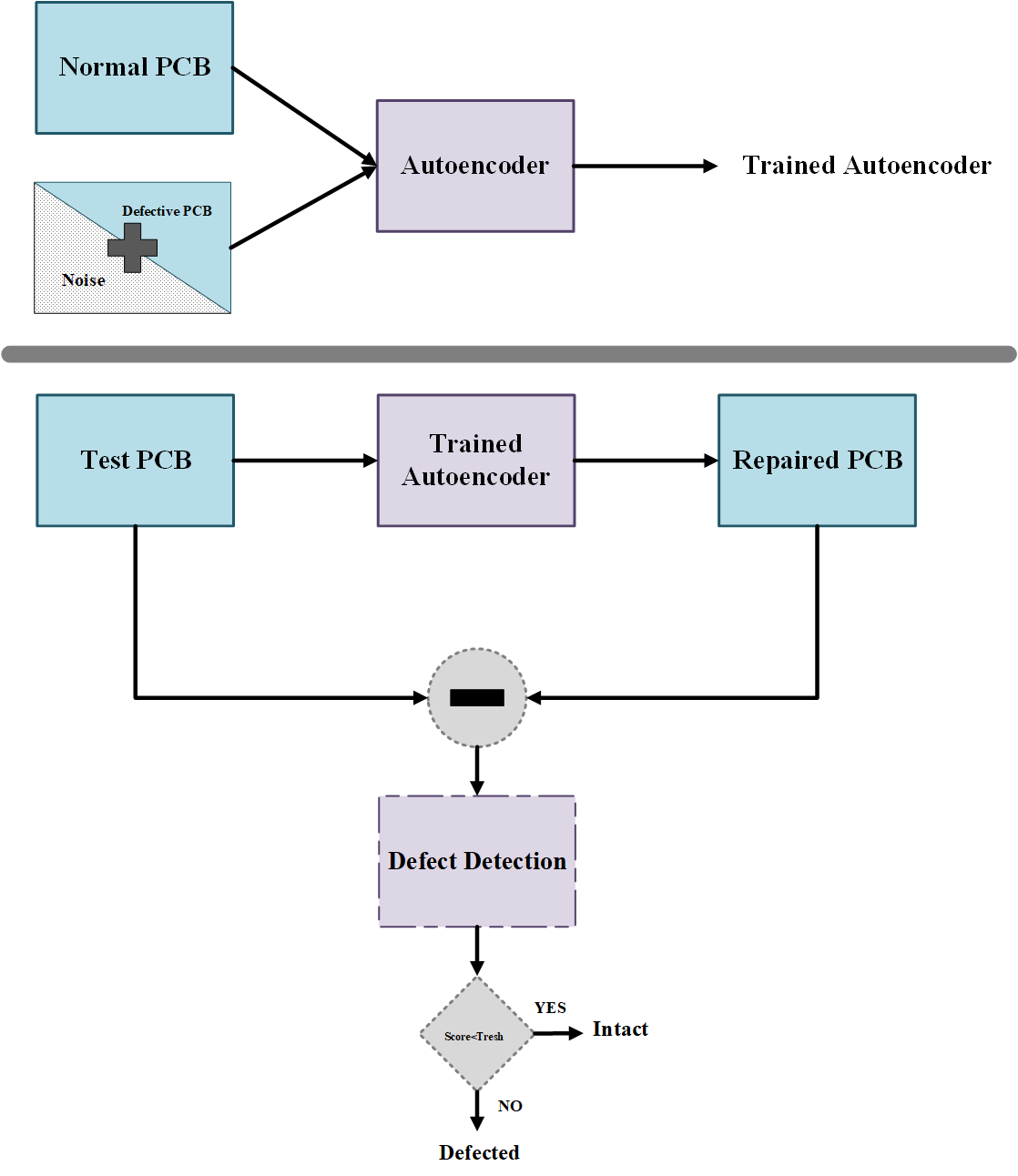}}
\caption{The overview of the system. a) salt-and-pepper noise is added to defective PCBs. b) Autoencoder is trained with pairs of normal PCBs and noisy PCBs. c) The output of the trained autoencoder is a repaired PCB. d) The defects are located with subtracting the output image from the input image.}
\label{fig}
\end{figure}

As it is shown in Figure 1, the proposed autoencoder is trained with a dataset containing image pairs of defective and intact PCBs. We added salt-and-pepper noise to the defective PCBs to force the autoencoder to learn better features. The output of the network is the repaired PCB. The difference rate between the input and the output determines whether the input PCB is defective or not. Moreover, by subtracting the repaired PCB from defective input, defective parts can be located. 

\subsection{Network Structure}
We have used denoising convolutional autoencoders in the proposed method to detect and repair defective PCBs. Denoising autoencoder is a special type of autoencoder which its inputs are noisy data, and it is forced to regenerate the intact samples by learning the features. Figure2 illustrates the designed autoencoder. 

\begin{figure}[htbpp]
\centerline{\includegraphics[width=\linewidth]{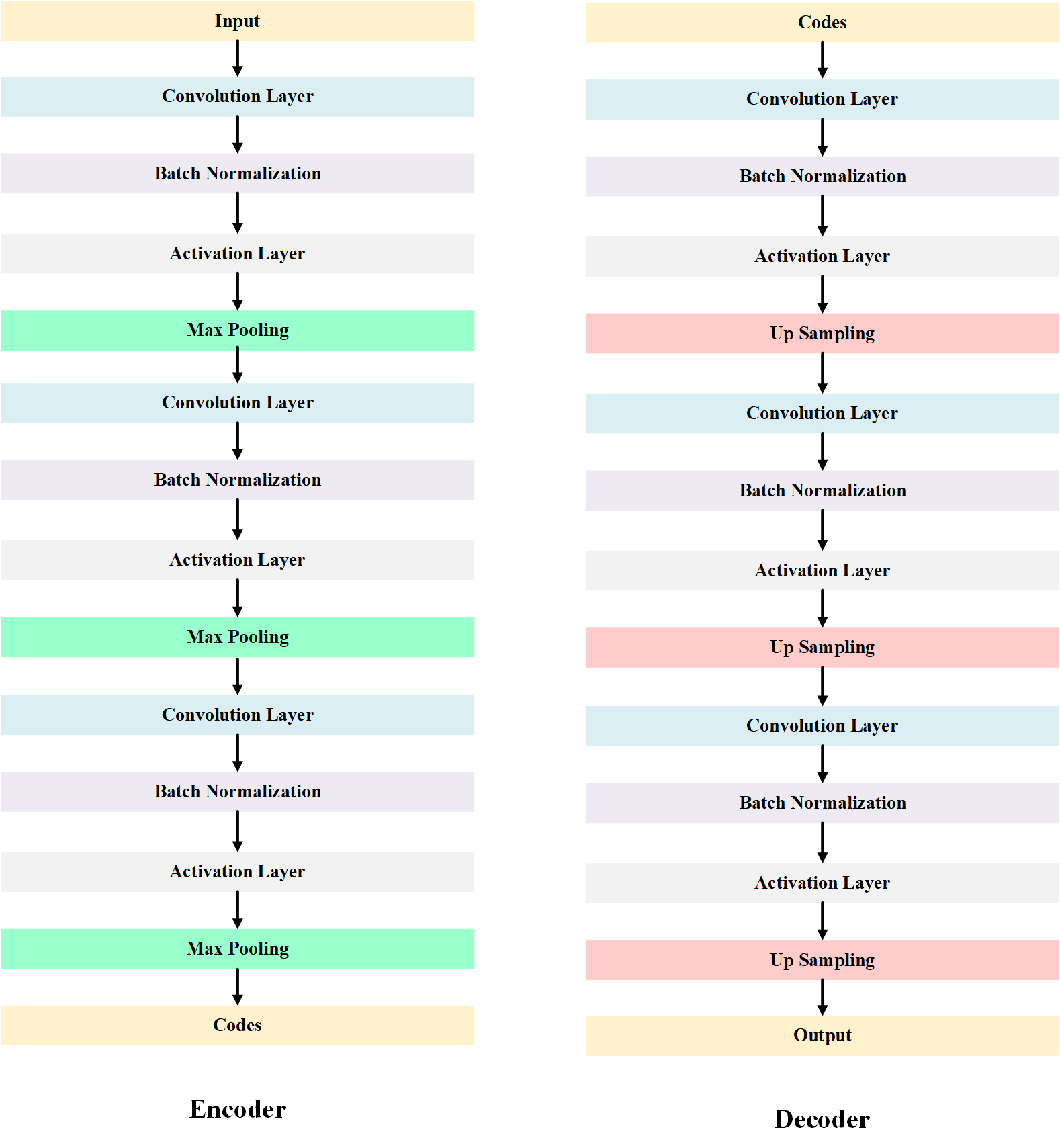}}
\caption{The structure of the proposed autoencoder. }
\label{fig}
\end{figure}

As it is shown in Figure 2, the proposed network has the following properties: 

\begin{itemize}
  \item The designed autoencoder consists of 2 parts, encoder, and decoder. Each part includes 3 layers. The encoder has 4 sublayers: convolution layer, batch normalization, activation layer and, Max pooling.  The decoder sublayers are convulsion layers, batch normalization, activation layer and, up sampling.
  \item The input images are 512*512 which are encoded into a vector contains 512*512 features. The decoder outputs a 512*512 image as the repaired image. The input and output of all layers are 512*512 vectors.
    \item The activation function used in the network is the Relu activation function. It is defined as:
  \[ Relu(x) = max(0,x) \]
  \item The loss function is the cross-entropy function which is defined as follows:    
	\[\frac{1}{N} \sum_{i=1}^{N}\left[y_{i} \log \left(\hat{y}_{i}\right)+\left(1-y_{i}\right) \log \left(1-\hat{y}_{i}\right)\right]\]
  \item To force the network to learn more useful features, salt-and-pepper noise is added to defective PCBs and they are used as autoencoder input. The network is forced to recover the repaired PCB from defective noisy images. In this way, it learns the main features of an intact PCB. 
  \item we have used transfer learning in order to achieve better results. At first, we trained an autoencoder to learn the features of normal PCBs. This trained model is used as the initial weights of our proposed network.

\end{itemize}

\subsection{Detecting Defects}
The trained network predicts the repaired PCB from the inputs. By subtracting the output from the input, the defective PCB is determined. The more the difference between the input and the output is, the more the PCB is defective.  Structural similarity (SSIM) is used to calculate the difference. We smooth the the result of SSIM to get more accurate difference and we draw contours of the resulted image. Each contour locates a defect and by computing the sum of these contours areas’, the final difference is calculated. Figure 3 shows an example of the defect detection procedure. 
 
\section{Experiment Results}
We have used the DeepPCB dataset that is provided by Tang et al \cite{b17} for our experiments. The dataset contains 1500 image pairs of a defective PCB with its aligned normal PCB. This dataset includes 6 kinds of defects. In this section, the proposed model is evaluated and the results are described.
The network is trained with image pairs of defective and normal PCBs. We added salt-and-pepper noise to the defective PCBs to achieve better results. Figure 3 shows a sample of the training images.

\begin{figure}[htbpp]
\centerline{\includegraphics[width=\linewidth]{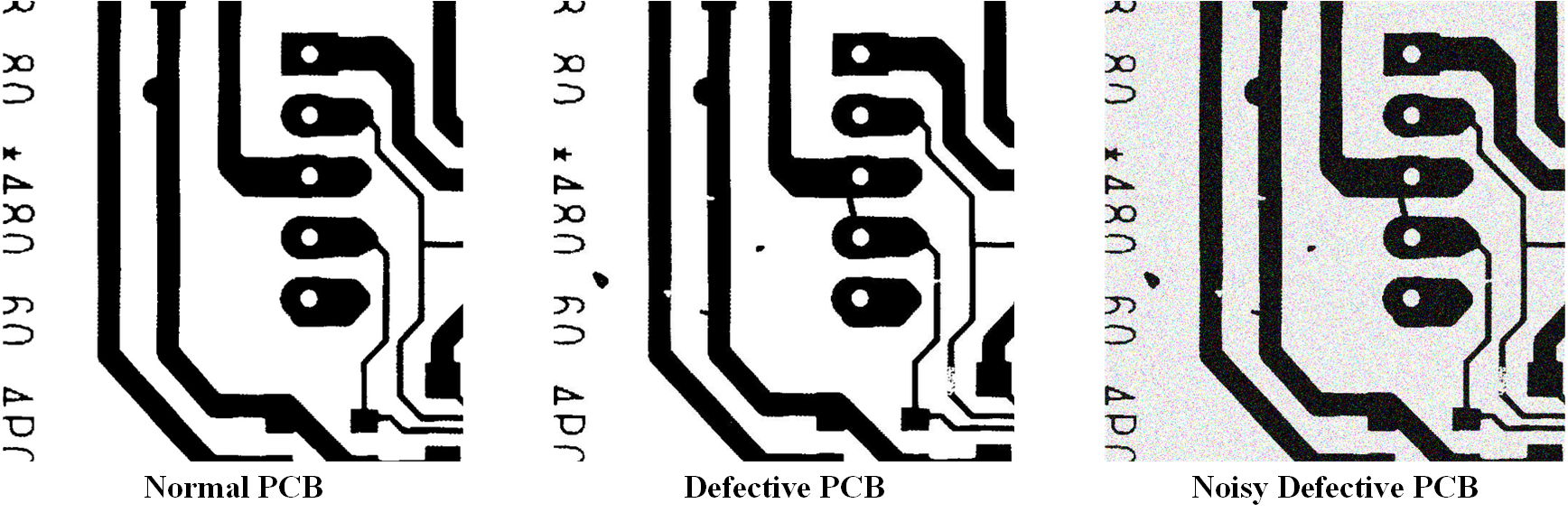}}
\caption{A sample of dataset which contains normal, defective and, created noisy PCBs. }
\label{fig}
\end{figure}

Before training the proposed denoising autoencoder, we first trained an autoencoder with only normal PCBs to learn their. Then we used the weights of that network for initial weights of our proposed network which increased the accuracy of the model. The validation loss of the network is 0.0867. Figure 4 shows the validation loss of these networks with batch size 2 over 17 and 4 epochs respectively.

\begin{figure}[htbpp]
\centerline{\includegraphics[width=\linewidth]{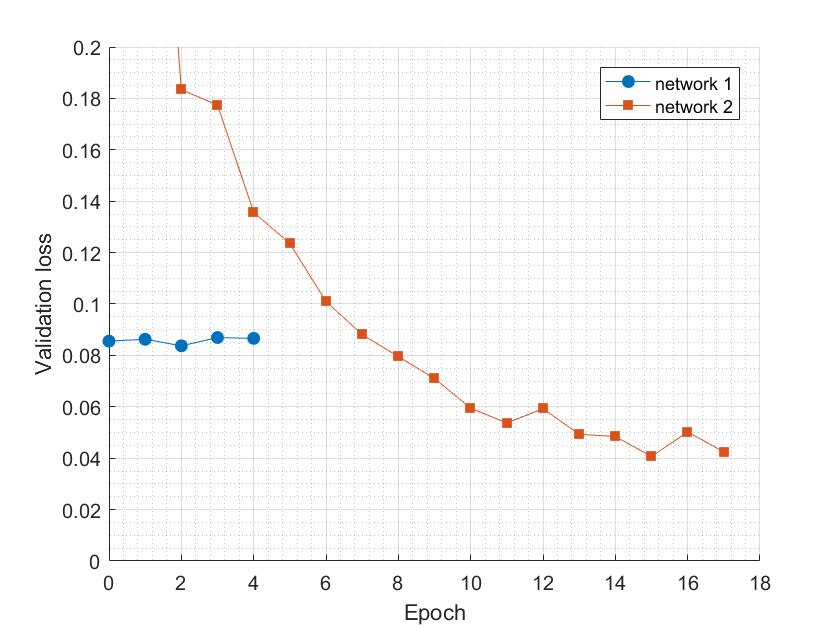}}
\caption{The validation loss of two networks. Network 2 indicates the validation loss for the first network that its weights are used in the proposed network and it is trained with only normal PCBs. Network 1 shows the validation loss for the proposed network which is a denoising autoencoder. }
\label{fig}
\end{figure}

The output of the trained network is a repaired PCB. The autoencoder does not change an intact PCB, so if the difference of the output and the input is small, the input is not defective. On the contrary, the difference between a defective PCB and the output is a significant amount. By subtracting the output from the input, the defective parts will be located. An example of the output of the system is shown in figure 5.

\begin{figure}[htbp]
	\subfigure[Defective PCBs]
    {
  \centerline{\includegraphics [width=\linewidth]{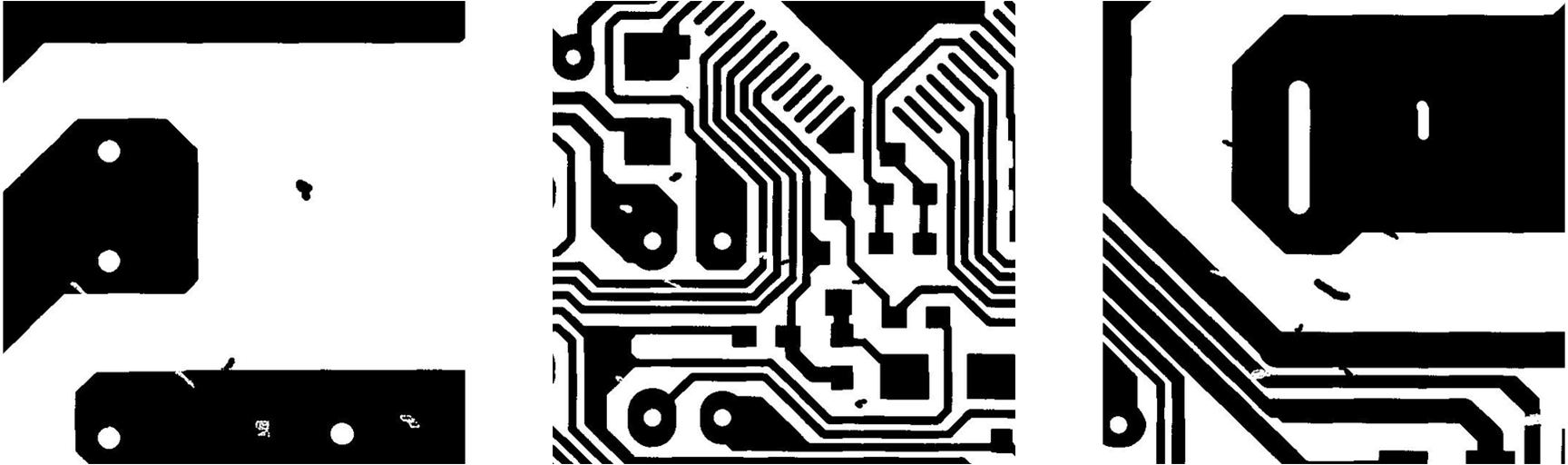}}
  }
  \subfigure[Repaired PCBs (outputs of the autoencoder)]
    {
  \centerline{\includegraphics[width=\linewidth]{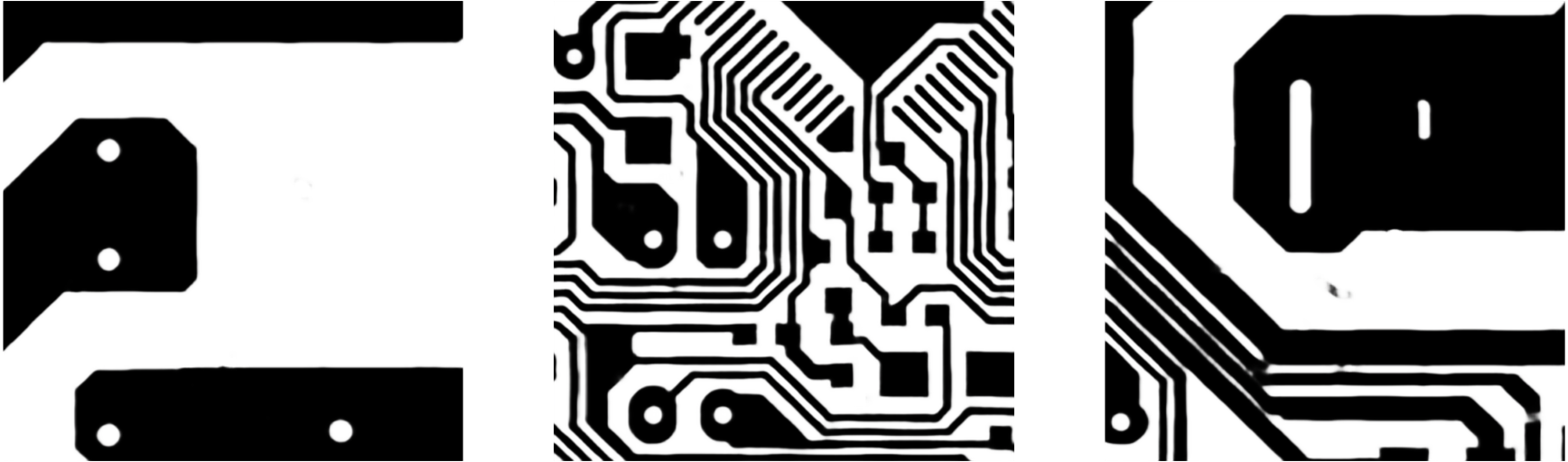}}
  }
  \subfigure[Located defects after subtract]
    {
  \centerline{\includegraphics[width=\linewidth]{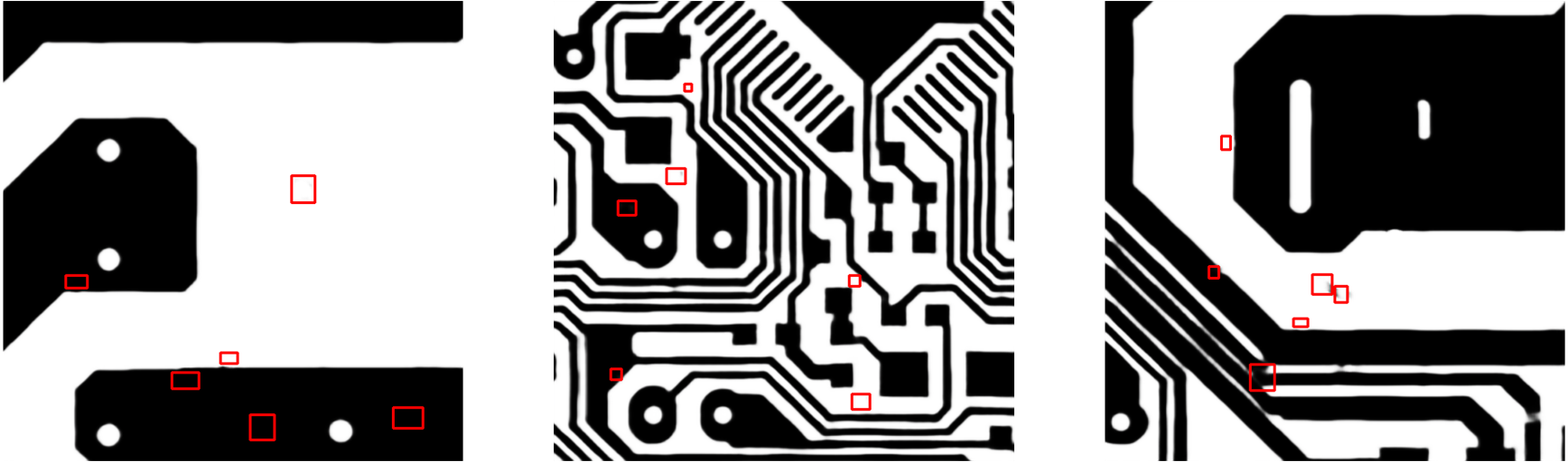}}
  }
\caption{examples of the output of the proposed model.}
\label{fig}
\end{figure}

The amount of difference between the input and the output determines whether the input is defective or not. If the difference is higher than a threshold, the input is considered as defective. Otherwise, it is considered as non-defective. To determine the best threshold, we calculate true positive rate recall, precision, selectivity, accuracy and, F-score for 4 different thresholds using following equations. Table 1 illustrates calculated amounts for these thresholds. 

\[recall=\frac{\mathrm{TP}}{\mathrm{P}}\]

\[precision=\frac{\mathrm{TP}}{\mathrm{TP}+\mathrm{FP}}=1-\mathrm{FDR}\]

\[selectivity=\frac{\mathrm{TN}}{\mathrm{N}}\]

\[accuracy=\frac{\mathrm{TP}+\mathrm{TN}}{\mathrm{P}+\mathrm{N}}\]

\[F-score=2 \times \frac{\text { precision } \times \text { recall }}{\text { precision }+\text { recall }}\]

\begin{table}
\centering
\caption{Comparing recall, precision, selectivity, accuracy and, F-score for 4 different thresholds}
\label{tab1}
\resizebox{\columnwidth}{!}{%
\begin{tabular}{|c|c|c|c|c|c|} 
\hline
\textbf{Threshold} & \textbf{recall} & \textbf{precision} & \textbf{selectivity} & \textbf{accuracy} & \textbf{F-score}  \\ 
\hline
50                 & 0.93            & 0.982              & 0.983                & 0.958             & 0.955             \\ 
\hline
100                & 0.97            & 0.983              & 0.983                & 0.975             & 0.976             \\ 
\hline
150                & 0.97            & 0.95               & 0.95                 & 0.958             & 0.959             \\ 
\hline
200                & 0.97            & 0.935              & 0.933                & 0.95              & 0.952             \\
\hline
\end{tabular}}
\end{table}

As it is shown in Table 1 when threshold equal to 100 the accuracy is maximum and it is equal to 0.975. It should be noted that this threshold works fine with all kinds of defects. 

\section{Conclusion and future work}
PCBs play a key role in producing electronic devices and the quality of the final product depends on its PCB. Therefore, the PCB should be flawless. In this paper, we proposed a defect detection method for PCBs based on denoising autoencoders. We trained the network with image pairs of the intact and defective PCBs. By learning the features of an intact PCB, our proposed method is able to repair the input and by subtracting the input from the output, the flaws are located. Our results proved the effectiveness of the proposed method. 

This paper presented a novel approach for defect detection in PCBs, however; the proposed method can be used to detect the defects in other kinds of products such as plastic injection molding products. Moreover, the subtracting algorithm can be improved to achieve more accurate results in locating the defects.

\end{document}